\documentclass[12pt,fleqn]{article}
\usepackage{amsmath}

\begin{document}

\title{\bf Singularity analysis of a spherical\\
Kadomtsev--Petviashvili equation}

\author{{\sc Ay\c{s}e Karasu-Kalkanli$^{1)}$,
Sergei~Yu.~Sakovich$^{2)}$}\\[16pt] \small{$^{1)}$Department of
Physics, Middle East Technical University,}\\[-6pt] \small{06531
Ankara, Turkey. E-mail: akarasu@metu.edu.tr}\\[6pt]
\small{$^{2)}$Institute of Physics, National Academy of
Sciences,}\\[-6pt] \small{220072 Minsk, Belarus. E-mail:
saks@tut.by}}

\date{}

\maketitle

\begin{abstract}
The (2+1)-dimensional spherical Kadomtsev--Petviashvili (SKP)
equation of J.-K.~Xue [Phys.\ Lett.\ A 314:479--483 (2003)] fails
the Painlev\'{e} test for integrability at the highest resonance,
where a nontrivial compatibility condition for recursion relations
appears. This compatibility condition, however, is sufficiently
weak and thus allows the SKP equation to possess an integrable
(1+1)-dimensional reduction, which is detected by the Weiss method
of truncated singular expansions.
\end{abstract}

\medskip

\section{Introduction}

Recently, Xue \cite{X} deduced a spherical Kadomtsev--Petviashvili
(SKP) equation for nonlinear dust acoustic waves in unmagnetized
dusty plasmas with effects of a nonplanar spherical geometry and a
transverse perturbation. This SKP equation can be brought into the
form
\begin{equation} \label{skp}
u_{xxxx} - 12 u u_{xx} - 12 u_x^2 + u_{xt} + \frac{1}{t^2} u_{yy}
+ \frac{1}{t} u_x + \frac{1}{y t^2} u_y = 0
\end{equation}
by rescaling its dependent and independent variables.

Xue \cite{X} noticed that the SKP equation \eqref{skp} admits two
exact reductions to (1+1)-dimensional equations: first, the
reduction $u=u(x,t)$ to the spherical Korteweg--de~Vries (SKdV)
equation $u_t + u_{xxx} - 12 u u_x + t^{-1} u = 0$, which is
believed to be a nonintegrable equation \cite{CD}; and second, the
reduction $u=u(z,t)$ with
\begin{equation} \label{z}
z = x - \tfrac{1}{4} y^2 t
\end{equation}
to the Korteweg--de~Vries (KdV) equation $u_t + u_{zzz} - 12 u u_z
= 0$. Having used this reduction to the integrable KdV equation,
Xue \cite{X} obtained an exact solitary wave solution of the SKP
equation.

Owing to the reduction to the KdV equation, the SKP equation
\eqref{skp} automatically possesses $N$-soliton solutions, with
any $N$, derivable from those of the KdV equation through the
variable $z$ \eqref{z}. Of course, the existence of such
(1+1)-dimensional solitons does not imply that the
(2+1)-dimensional SKP equation is integrable. Moreover, the
existence of the reduction to the SKdV equation suggests that the
SKP equation is nonintegrable.

In the present paper, we study the integrability of the SKP
equation \eqref{skp} directly, not using its reductions. In
Section~\ref{s2}, we show that the SKP equation does not pass the
Painlev\'{e} test for integrability due to some nondominant
logarithmic branching of its general solution. The singularity
analysis indicates, however, that many special solutions of the
SKP equation are free from this logarithmic branching. In order to
select those single-valued special solutions, we use the Weiss
method of truncated singular expansions in Section~\ref{s3}, and
in this way surprisingly obtain the reduction of the SKP equation
to the KdV equation. Section~\ref{s4} contains concluding remarks.

\section{Singularity analysis} \label{s2}

Let us show that the SKP equation \eqref{skp} does not pass the
Painlev\'{e} test for integrability. We follow the so-called
Weiss--Kruskal algorithm of singularity analysis \cite{WTC,JKM}.

Starting the singularity analysis, we substitute $u = u_0 (y,t)
\phi^{\alpha} + \dotsb + u_r (y,t) \phi^{r+\alpha} + \dotsb $ with
$\partial_x \phi (x,y,t) = 1$ into the SKP equation \eqref{skp},
and find that the singular behavior of a solution $u$ corresponds
to $\alpha = -2$ with $u_0 = 1$, the positions $r$ of resonances
being $r = -1, 4, 5, 6$. This is the generic branch representing
the general solution.

Then, assuming that the singular behavior of $u$ near a
hypersurface $\phi (x,y,t) = 0$ with $\phi_x = 1$ is determined by
the expansion
\begin{equation} \label{ex}
u = \sum_{n=0}^{\infty} u_n (y,t) \phi^{n-2} ,
\end{equation}
we obtain from \eqref{skp} the following recursion relations for
the coefficients $u_n$:
\begin{multline} \label{rr}
(n-2)(n-3)(n-4)(n-5) u_n\\ + (n-4)(n-5) \left( - 6 \sum_{i=0}^n
u_i u_{n-i} + \left( \phi_t + \frac{1}{t^2} \phi_y^2 \right)
u_{n-2} \right)\\ + (n-5) \Biggl( \left( \frac{1}{t} +
\frac{1}{t^2} \phi_{yy} + \frac{1}{y t^2} \phi_y \right) u_{n-3} +
\left( 1 + \frac{2}{t^2} \phi_y \right) \partial_t u_{n-3}
\Biggr)\\ + \frac{1}{t^2} \partial_y^2 u_{n-4} + \frac{1}{y t^2}
\partial_y u_{n-4} = 0, \qquad n=0,1,2, \dotsc ,
\end{multline}
where $u_{-4} = u_{-3} = u_{-2} = u_{-1} = 0$ formally.

At $n=0,1,2,3$, the recursion relations \eqref{rr} give us,
respectively,
\begin{align}
u_0 & = 1 ,\\ u_1 & = 0 ,\\ u_2 & = \frac{1}{12} \left( \phi_t +
\frac{1}{t^2} \phi_y^2 \right) ,\\ u_3 & = - \frac{1}{12} \left(
\frac{1}{t} + \frac{1}{t^2} \phi_{yy} + \frac{1}{y t^2} \phi_y
\right) .
\end{align}
At the resonances $n=4$ and $n=5$, where the coefficients $u_4
(y,t)$ and $u_5 (y,t)$ are not determined, the recursion relations
\eqref{rr} turn out to be compatible. However at the highest
resonance, $n=6$, where the coefficient $u_6 (y,t)$ is not
determined, we obtain from \eqref{rr} the following nontrivial
compatibility condition:
\begin{equation} \label{cc}
\left( \phi_y + \frac{1}{2} y t \right) \left( \phi_{yy} +
\frac{1}{2} t \right) = 0 ,
\end{equation}
which means that we should modify the expansion \eqref{ex} by
introducing additional logarithmic terms, starting from the one
proportional to $\phi^4 \log \phi$.

Consequently, the SKP equation \eqref{skp} does not pass the
Painlev\'{e} test for integrability due to the nondominant
logarithmic branching of its solutions. The observed analytic
properties of the SKP equation suggest that it cannot possess any
good Lax pair.

There is an interesting conjecture, formulated by Weiss \cite{Wc},
that the differential constraints, which arise in the singularity
analysis of nonintegrable equations, are always integrable
themselves (see \cite{S} for further discussion on this
conjecture). In the present case of the SKP equation \eqref{skp},
we find that, in accordance with the Weiss conjecture, the
compatibility condition \eqref{cc} with $\phi_x = 1$ can be solved
exactly, the result being
\begin{equation} \label{phi}
\phi = x - \frac{1}{4} y^2 t + y f(t) + g(t)
\end{equation}
with any $f(t)$ and $g(t)$.

We see from \eqref{phi} that the compatibility condition
\eqref{cc} is not very restrictive. The class of single-valued
solutions of the SKP equation \eqref{skp}, which are free from the
nondominant logarithmic branching, is wide: it is determined by
the Laurent type expansion \eqref{ex} containing three arbitrary
functions of two variables and two arbitrary functions of one
variable, namely, $u_4 (y,t)$, $u_5 (y,t)$, $u_6 (y,t)$, $f(t)$
and $g(t)$. For this reason, one can hope to find many special
single-valued solutions of the SKP equation \eqref{skp} in some
closed form, whereas the existing techniques provide no closed
expressions for solutions possessing nondominant logarithmic
singularities.

\section{Truncated expansion} \label{s3}

Let us apply the method of truncated singular expansions of Weiss
\cite{Wt} to the SKP equation \eqref{skp}. This method, which is
able to produce B\"{a}cklund transformations and Lax pairs for
integrable nonlinear systems, is also useful in nonintegrable
cases for finding explicit special solutions \cite{CT}.

Substituting the truncated singular expansion
\begin{equation} \label{tr}
u = \frac{u_0 (x,y,t)}{\phi (x,y,t) ^2} + \frac{u_1 (x,y,t)}{\phi
(x,y,t)} + u_2 (x,y,t)
\end{equation}
into the SKP equation \eqref{skp} and collecting terms with equal
degrees of $\phi$, we obtain the following:
\begin{align}
u_0 & = \phi_x^2 , \label{u0} \\ u_1 & = -\phi_{xx} , \label{u1}
\\ u_2 & = \frac{\phi_{xxx}}{3 \phi_x} - \frac{\phi_{xx}^2}{4
\phi_x^2} + \frac{\phi_y^2}{12 t^2 \phi_x^2} + \frac{\phi_t}{12
\phi_x}, \label{u2}
\end{align}
\begin{multline} \label{e1}
\phi_{xxxx} - \frac{4 \phi_{xx} \phi_{xxx}}{\phi_x} + \frac{3
\phi_{xx}^3}{\phi_{x}^2} - \frac{\phi_y^2 \phi_{xx}}{t^2 \phi_x^2}
- \frac{\phi_t \phi_{xx}}{\phi_x} \\ + \phi_{xt} +
\frac{\phi_{yy}}{t^2} + \frac{\phi_x}{t} + \frac{\phi_y}{y t^2} =
0 ,
\end{multline}
\begin{equation} \label{e2}
\left( y t \phi_x + 2 \phi_y \right) \left( \phi_y^2 \phi_{xx} - 2
\phi_x \phi_y \phi_{xy} + \phi_x^2 \phi_{yy} + \tfrac{1}{2} t
\phi_x^3 \right) = 0 .
\end{equation}
We see that, for any solution $\phi (x,y,t)$ of the overdetermined
nonlinear system \eqref{e1}--\eqref{e2}, the truncated singular
expansion \eqref{tr} with the coefficients given by
\eqref{u0}--\eqref{u2} generates a solution $u(x,y,t)$ of the SKP
equation \eqref{skp}. However, in order to use this fact, we need
to solve the system \eqref{e1}--\eqref{e2}.

First, we solve the equation \eqref{e2}, which is equivalent to
\begin{equation} \label{eq}
\phi_y^2 \phi_{xx} - 2 \phi_x \phi_y \phi_{xy} + \phi_x^2
\phi_{yy} + \tfrac{1}{2} t \phi_x^3 = 0
\end{equation}
because any solution of $y t \phi_x + 2 \phi_y = 0$ satisfies
\eqref{eq}. Setting $x$ to be the new dependent variable,
\begin{equation} \label{ch}
x = \psi (\phi ,y,t) ,
\end{equation}
we rewrite \eqref{eq} in the linear form
\begin{equation} \label{le}
\psi_{yy} = \tfrac{1}{2} t .
\end{equation}
The general solution of \eqref{le} and the expression \eqref{ch}
give us the following implicit general solution of the equation
\eqref{e2}:
\begin{equation} \label{is}
x = \tfrac{1}{4} y^2 t + a(\phi ,t) y + b(\phi ,t) ,
\end{equation}
where $a$ and $b$ are arbitrary functions.

Now, having solved the equation \eqref{e2}, we use \eqref{is} and
find that the equation \eqref{e1} is equivalent to
\begin{gather}
a = 0 , \label{a} \\ b_{\phi t} + \frac{b_{\phi \phi \phi
\phi}}{b_{\phi}^3} - \frac{6 b_{\phi \phi} b_{\phi \phi
\phi}}{b_{\phi}^4} + \frac{6 b_{\phi \phi}^3}{b_{\phi}^5} = 0 .
\label{b}
\end{gather}
Owing to the condition \eqref{a}, we get from \eqref{is} that
$\phi$ is in fact a function of two variables,
\begin{equation} \label{om}
\phi (x,y,t) = \omega (z,t) ,
\end{equation}
where $z$ is given by \eqref{z}. Then the condition \eqref{b} is
equivalent to
\begin{equation} \label{oe}
\left( \frac{\omega_t}{\omega_z} + \frac{\omega_{zzz}}{\omega_z} -
\frac{3 \omega_{zz}^2}{2 \omega_z^2} \right)_z = 0 .
\end{equation}

We have solved the overdetermined nonlinear system
\eqref{e1}--\eqref{e2}: its general solution is \eqref{om}, where
$z$ is given by \eqref{z} and $\omega$ is any solution of
\eqref{oe}. Using this, we find from \eqref{tr} with
\eqref{u0}--\eqref{u2} that the most general solution $u$,
obtainable for the SKP equation \eqref{skp} by the method of
truncated singular expansions, is
\begin{equation} \label{u}
u(x,y,t) = v(z,t)
\end{equation}
with $z$ given by \eqref{z} and any function $v$ satisfying the
equation
\begin{equation} \label{red}
\left( v_t + v_{zzz} - 12 v v_z \right)_z = 0.
\end{equation}
Consequently, the Weiss method \cite{Wt}, being applied to the SKP
equation, rediscovers the reduction of this (2+1)-dimensional
nonintegrable equation to the (1+1)-dimensional integrable KdV
equation.

\section{Conclusion} \label{s4}

Let us summarize the obtained results. The discovered analytic
properties of the SKP equation \eqref{skp} suggest that this
equation cannot admit any good Lax pair but can possess many
single-valued solutions. The attempt of selecting those
single-valued solutions by using the truncated singular expansion
leads not to a class of explicit special solutions but to an exact
reduction of the studied nonintegrable equation to a lower
dimensional integrable equation (as far as we know, this
phenomenon is observed for the first time). It is very likely,
however, that the truncated singular expansion represents not all
single-valued solutions of the SKP equation \eqref{skp}. Indeed,
positions $\phi = 0$ of singularities of a generic single-valued
solution are determined by \eqref{phi} with arbitrary $f(t)$ and
$g(t)$, whereas positions of singularities of solutions
\eqref{u}--\eqref{red} are restricted by $f=0$. Therefore it would
be interesting to find any closed form solution of the SKP
equation \eqref{skp} without this restriction $f=0$ imposed on
positions of its singularities.

\end{document}